\documentclass[twocolumn, switch]{article} 

\usepackage{arxiv}
\usepackage{caption}
\usepackage{hyperref}       
\usepackage{url}            
\usepackage{booktabs}       
\usepackage{amsfonts}       
\usepackage{nicefrac}       
\usepackage{microtype}      
\usepackage{lipsum}

\usepackage{amssymb}
\usepackage{latexsym}

\usepackage{comment}
\usepackage{graphicx}

\usepackage{natbib}
\setcitestyle{square,numbers,comma}

\title{A Framework for Challenge Design: \\ Insight and Deployment Challenges to Address Medical Image Analysis Problems}

\author{
  Adri\"enne M. Mendrik \\
  Netherlands eScience Center\\
  Science Park 140\\
  1098 XG Amsterdam, the Netherlands \\
  \texttt{a.m.mendrik@gmail.com} \\
   \And
 Stephen R. Aylward\\
  Kitware, Inc.\\
  101 East Weaver Street, Suite G4\\
  Carrboro, North Carolina, 27510, USA \\
  \texttt{stephen.aylward@kitware.com} \\
}

\begin{document}
\twocolumn[ 
  \begin{@twocolumnfalse} 

\maketitle

\begin{abstract}
In this paper we aim to refine the concept of grand challenges in medical image analysis, based on statistical principles from quantitative and qualitative experimental research. We identify two types of challenges based on their generalization objective: 1) a deployment challenge and 2) an insight challenge. A deployment challenge's generalization objective is to find algorithms that solve a medical image analysis problem, which thereby requires the use of a quantitative experimental design. An insight challenge's generalization objective is to gain a broad understanding of what class of algorithms might be effective for a class of medical image analysis problems, in which case a qualitative experimental design is sufficient. Both challenge types are valuable, but problems arise when a challenge's design and objective are inconsistent, as is often the case when a challenge does not carefully consider these concepts. Therefore, in this paper, we propose a theoretical framework, based on statistical principles, to guide researchers in challenge design, documentation, and assessment. Experimental results are given that explore the factors that effect the practical implementation of this theoretical framework.
\end{abstract}
\vspace{0.35cm}


\vspace{0.5cm}

\end{@twocolumnfalse} 
] 

\section{Introduction}
\label{sec1}
A challenge is an online competition that uses data, truth and metrics to evaluate the performance of automatic algorithms, submitted by participants, with respect to a research problem. In the medical image analysis field, close to 150 challenges have been organized thus far \citep{Ginneken18}. 

Well designed challenges have the potential to unite and focus the field on creating solutions to important problems as well as generating new insights and identifying common themes for future research. Several breakthrough technologies can attribute the recognition of their capabilities to challenges. The ImageNet challenge revealed the utility of neural networks \citep{Russakovsky2015}, and in the medical image analysis field the Camelyon Challenge, for example, showed that deep learning algorithms can, under certain conditions, outperform a panel of 11 pathologists in detecting lymph node metastases on whole-slide pathology images \citep{bejnordi2017diagnostic}. Furthermore, the indirect impact of challenges should not be underestimated. For example, the Retrospective Image Registration Experiment challenge revealed the utility of mutual information \citep{West97comparison}, and the NetFlix challenge boosted the field of recommender systems \citep{gomez16netflix,Netflix17}. From the field of robotics it is also evident that competitions such as the RoboCup \citep{kitano97robocup} can be a catalyst for innovation \citep{hart18,pavez18} as well as interdisciplinary teamwork.  

Poorly designed and poorly documented challenges, however, can equally impede a field, e.g. lead to false performance expectations and promote unfruitful research directions. The potential detriment of challenges has inspired a critical assessment of the proposal and review processes for challenges at the international Medical Image Computer and Computer-Assisted Intervention (MICCAI) conference and has led to a definition of best practices for challenge design \citep{Reinke18, lena18}. The main criticism raised in that assessment is based on the observation that small changes to the underlying challenge data have the potential to drastically change the rank order in the leaderboard. Our framework emphasizes that this criticism is exacerbated by the practice of leaderboard climbing \citep{Hardt15,whitehill17}, which is characterized by participants focusing on incrementally improving metric results rather than gaining new insights or solving the challenge's driving problem. 

Challenge leaderboards are valuable, but it is critical to understand the scope of a challenge when interpreting its leaderboard. Leaderboards do provide a quick overview of algorithm performance and can reveal trends and boost collaboration as well as competition; however, they are also constrained by the design of the underlying data, truth and metrics. These design constraints limit the conclusions that \emph{can} be drawn from a challenge, i.e., a challenge's design limits its scope. The theoretical framework for challenge design that we are proposing can be used to clearly describe a challenge's scope, i.e., what conclusions can be and are resolved by a challenge. Thus, the aim of the framework is twofold: first, to guide challenge organizers in challenge design and description, and second, to clarify the scope of a challenge for challenge participants and the community at large. The latter is critical in order to moving beyond a challenge's leaderboard, i.e., to understanding the relevance of a challenge and identifying commonalities and differences across multiple challenges.

We propose that to understand and define the scope of a challenge, it is critical to distinguish two fundamental types of challenges: 1) an \textbf{insight challenge} that can be used to gain insight into a research problem by analyzing the performance of multiple algorithms, and 2) a \textbf{deployment challenge} that can be used to test which algorithms successfully solve a research problem. A deployment challenge follows a quantitative research study design, whereas an insight challenge follows a qualitative design. Our work borrows heavily from research study design practices that are well described in other fields, such as educational research \citep{Creswell02}.  We translate those practices into a framework for challenge design by viewing challenges as research studies on algorithm performance for a specific problem, as described next.

\section{Insight and Deployment Challenges}
\label{sec2}
\begin{figure}
\centering
\includegraphics[scale=.3]{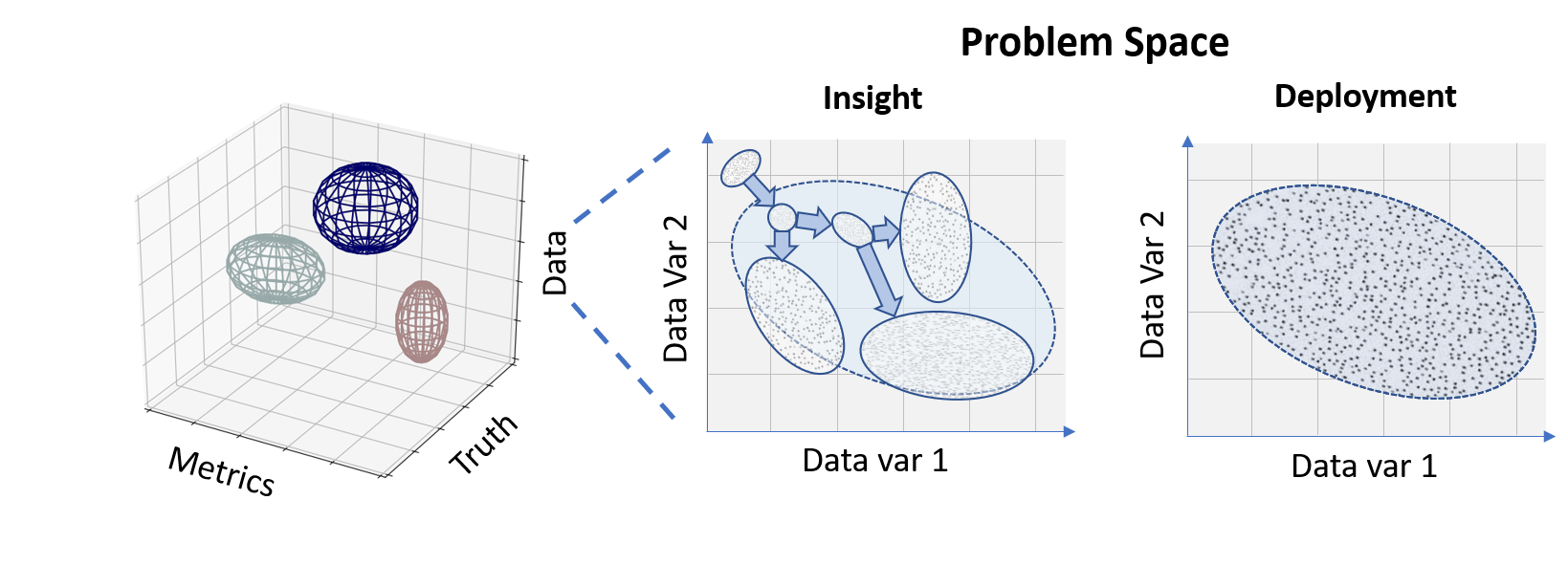}
\caption{\label{fig1}Challenges can be conceptualized as statistical samples from a high dimensional space defined by the data, truth, and metrics of that challenge. \textbf{Left}: Three challenges are illustrated as ellipses (e.g., iso-probability contours) in a data, truth, and metric space. \textbf{Middle}: The individual ellipses represent Insight challenges that systematically explore regions within a problem space (i.e., use purposive sampling). \textbf{Right}: The span of a problem space must be completely covered by a deployment challenge (i.e., use representative sample), for that challenge to determine whether a problem has been solved. The Middle and Right graphs emphasize how the data used in challenges must cover the problem space, but as illustrated in the graph on the Left, the metrics and truth used in challenges must also cover (i.e., be representative of) the relevant problem space.}
\end{figure}

When challenges are conceptualized as studies of algorithm performance that address a specific problem, the complexity of defining and sampling the relevant problem space is revealed. Examples of research problems in medical image analysis include tissue segmentation, image registration, disease classification, pathology detection, and so on. Clearly, any given challenge will only cover a small portion of the space associated with any of those problems. That is, the scope of a challenge is constrained by its underlying data, truth and metrics, i.e., by its study design. This is illustrated in Figure \ref{fig1}.

As mentioned in the previous section, we propose to categorize challenges into two fundamental types that can be used to address a research problem: 1) an insight challenge that is based on a qualitative study design, and 2) a deployment challenge that is based on a quantitative study design. This categorization is driven by the two primary types of research study designs, that are distinguished by their use of qualitative or quantitative experiments \citep{Onwuegbuzie07, Creswell02}. The main difference between a qualitative and a quantitative study design is whether or not the results of the challenge can be generalized to the statistical population of the research problem. In other words, whether the submitted algorithms can be expected to perform similarly on other data from the research problem, beyond the test data provided by the challenge.

Although a challenge in itself is a quantitative system of comparison, most challenges that have been organized in the medical image analysis field thus far follow a qualitative experimental design in consideration of their underlying data, truth or metrics. For example, obtaining a randomized, representative set of data with gold standard truth for a medical image analysis problem for a quantitative experiment is often expensive, labor intensive, time consuming, and restricted by ethical concerns. Therefore most challenges incorporate indirect metrics, bronze standards for truth \citep{Jannin02}, and/or non-randomly selected datasets, which are the hallmarks of a qualitative experiment. As qualitative experiments, their leaderboards or subsequent tests cannot be used to draw quantitative conclusions about the research problem targeted by the challenge. Attempting to compute quantitative statistics from qualitative experiments is analogous to applying a t-test when the source data is not randomly sampled or does not have a normal distribution. This is also evident from the work by Reinke et al. \citep{Reinke18}, that can be interpreted to show that a quantitative system should be used with caution for challenges.

\begin{figure*}
\centering
\includegraphics[scale=.75]{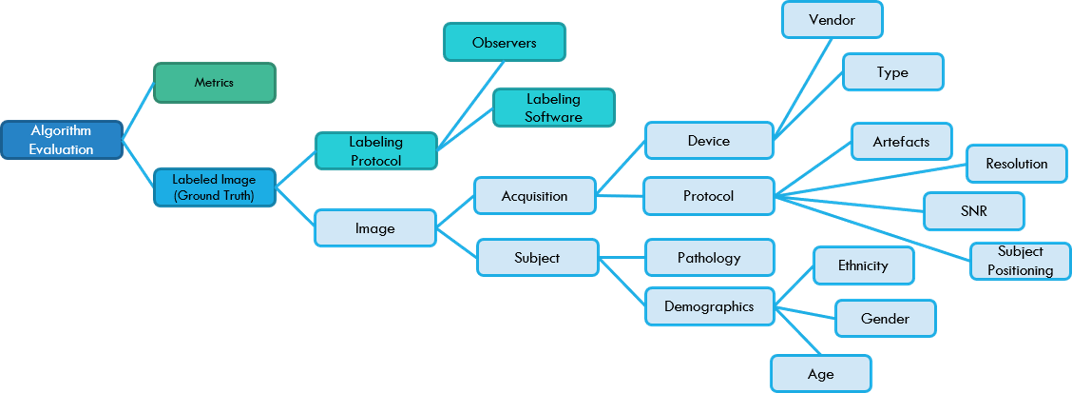}
\caption{\label{fig2}Example of potential statistical population variables in a medical image analysis problem that can influence algorithm performance, e.g., scanner protocols as well as a patient's age can influence automated brain tissue segmentation in MRI. The relevant variables define the domain of the statistical population that is relevant to a research problem. Note that not only the data, but also the truth and metrics are part of the statistical population of a challenge.}
\end{figure*}

The existence of constrains on the quantitative conclusions that can be drawn from qualitative challenges, however, does not mean that the leaderboards of qualitative challenges are not valuable. Whereas quantitative study design is focused on testing a hypothesis on a representative sample of the statistical population to ensure generalization, qualitative design is aimed at (1) discovering potential hypotheses and influential population variables from specific non-randomly sampled data and (2) uncovering the underlying reasoning \citep{Foley2015, Hanson2011, Kaplan2005}. Qualitative and quantitative study design can strengthen each other \citep{Shah06, brannen2017combining} and work together towards solving a research problem. Insight challenges, in particular, can be used to draw conclusions that lead to new insights with respect to algorithm performance for a certain research problem, when (1) the design choices for those insight challenges are openly described with respect to data sampling, ground truth protocol and performance metrics used; and when (2) those challenges clearly follow from a research problem, purpose statement and research question. Analysis of the challenge results, in combination with knowledge of the algorithms' methodologies, can lead to the discovery of new theories and research directions. The systematic analysis of qualitative data is commonly used in other fields, such as social sciences. An example of systematic qualitative analysis is the grounded theory approach \citep{corbin1990grounded,Glaser68}.

Insights from challenges can in turn inspire new challenges that address the same research problem from a different angle, e.g., that address a particularly interesting or difficult portion of the broader research problem. This is illustrated in Figure \ref{fig1}, in which insight challenges occupy a limited part of the problem space that has three meta-axes: data, truth, and metrics. In reality, the problem space (statistical population) of a challenge (algorithm evaluation) is much more complex. A problem space is typically an extremely high-dimensional space that incorporates a multitude of statistical population variables.

Figure \ref{fig2} provides an example of potential statistical population variables associated with an algorithm evaluation problem space in medical image analysis, such as brain tissue segmentation in MRI. Algorithm performance can be influenced by variations in data, truth, and metrics. In terms of data, in this case the image, variations in image acquisition and the subject imaged can have a large influence on algorithm performance. Additionally, the truth that is provided for both training and testing the algorithm adds another layer of variation, in terms of labeling protocol (labeling software, manual / semi-automatic labeling) and observers (inter- and intra-observer variability). Insight challenges can provide insight into how algorithms are influenced by these variables in order to determine which variables should be taken into account when designing a deployment challenge. For example, some methods might be robust against certain types of pathology, while others are not. Furthermore, the metrics that are used to measure the distance between an algorithm's results and the provided truth must also be taken into account. Until there is consensus on one metric or a combination of metrics that work best for evaluating algorithms for a specific problem, there are a variety of metrics and ranking schemes that can be used or explored via insight challenges. 

\section{Challenge Framework}\label{sec3}
\begin{figure*}
\centering
\includegraphics[scale=1.0]{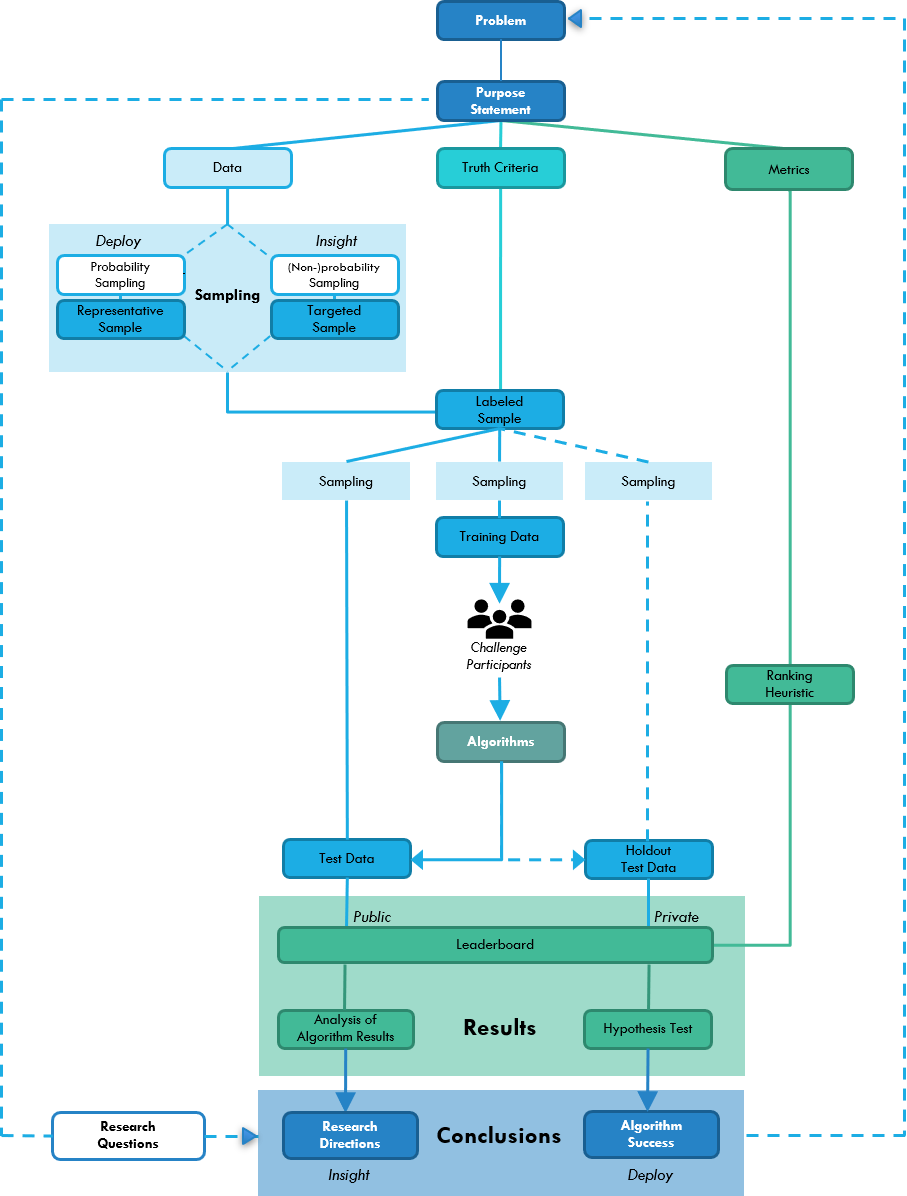}
\caption{\label{fig3}Framework for challenge design.}
\end{figure*}

Figure \ref{fig3} provides a schematic overview of the proposed framework. The design of a challenge starts with a description of the \textbf{problem} that it aims to address. As described in Section \ref{sec2}, this problem has a certain problem space (Figure \ref{fig1}), which encompasses the variations that are expected to impact algorithm performance (example Figure \ref{fig2}). 

The \textbf{purpose statement} describes which part of the problem is going to be the focus of the challenge, i.e., which part of the problem space is covered by the challenge. As part of describing the problem space and purpose of a challenge, it is good practice to define which \textbf{research questions} the challenge aims to answers \cite{Creswell02}. Based on the purpose statement and the challenge type (insight or deployment), the \textbf{data}, \textbf{truth criteria}, and \textbf{metrics} are then defined. 

The truth criteria, determine the task to be addressed by the challenge, given the data. The combination of truth criteria and data provides the \textbf{labeled sample} that can be used to train algorithms for a challenge's specific task and to test their performance. Division of the labeled samples into training and test data should follow the sampling scheme of either an insight or a deployment design (Sec \ref{sec31},\ref{sec32}). In both designs, the \textbf{training data} should be representative of the \textbf{test data}, and in case of a hold-out set, the \textbf{hold-out data} for a challenge's private leaderboard should be representative of the test data used for the public leaderboard.

A challenge's metrics determine the distance between the defined truth and the output of the \textbf{algorithms} submitted by \textbf{challenge participants} on the test data. A \textbf{ranking heuristic} can combine various metrics into a single value that can be used for ranking algorithms for the \textbf{leaderboard}. The ranking heuristic should be chosen so that it reflects the performance of algorithms with respect to the purpose statement. In this way, well-designed leaderboards can be of great value by providing quick insights into algorithm performance. 

As mentioned in Section \ref{sec2}, while leaderboards can be of great value, leaderboards should be interpreted with caution, e.g., with respect to their underlying data, truth criteria, and metrics. Sections \ref{sec31} and \ref{sec32} describe the interpretation of the leaderboard for a deployment or insight design. In general, leaderboards of insight challenges should be assessed as rough estimates of algorithm performance with respect to the challenge purpose, after which further \textbf{analysis of algorithm results} is required to gain insight. The leaderboard of a deployment challenge should be more precise allowing for \textbf{hypothesis testing} to check whether an algorithm solves the challenge's problem. 

One important consideration in leaderboard design is the theory of static versus adaptive data analysis \cite{Hardt15dec}. Static data analysis is the more traditional view that the test data can only be used once, in order to test generalization. This prevents leaking knowledge from the test data into algorithm design. In challenge design, static data analysis can be used for deployment challenges to ensure generalization and allow for testing whether a problem is solved. A hold-out test dataset can be selected that is sufficiently large and representative of the test data that is used for the public leaderboard. The disadvantage of this approach is that it requires significantly more data. Approaches like the re-usable hold-out \cite{Dwork15, Hardt15Google} offer a potential solution to enable data re-use without exhausting the dataset, i.e., to preserve the opportunity to test for generalization. This approach offers adaptive data analysis while reducing the risk of overfitting on the test data. There are two forms of adaptivity \cite{Hardt15dec}: 1) Algorithmic adaptivity and 2) Human adaptivity. 

The practice of leaderboard climbing \cite{Hardt15,whitehill17} is a form of algorithmic adaptivity. It is characterized by participants focusing on incrementally improving ranking results on the public leaderboard rather than advancing science or solving the challenge’s driving problem. A potential way to combat this is to only update the leaderboard in case of a significant change \cite{Blum15}, or use the method behind differential privacy \cite{dwork2011}, similar to what is done with data for the re-usable hold-out, to add noise to the evaluation scores. At its core, differential privacy is a notion of stability requiring that any single sample should not influence the outcome of the analysis significantly \cite{Hardt15Google}. Another option is to limit the number of allowed resubmissions, but this is harder to check and easier to circumvent by participants. 

Human adaptivity is more complex to quantify. If humans work adaptively with data, they gain insight into the data \cite{Hardt15dec}, and certain assumption can be made about the data that are taken into account in algorithm development. This could be beneficial, especially since humans are much better able to generalize than machine learning algorithms, but can affect algorithm validation. In terms of challenge design, hidden test data would aid in combating human adaptivity on the test data. This can be done by offering the opportunity to submit algorithms that are then run on the test data by the challenge organizers or by an automated challenge platform. Further research is required to determine an effective balance between avoiding human and algorithmic adaptivity and offering the opportunity for further analysis of algorithm results for insight challenges. 

\subsection{Deployment Design}\label{sec31}
For a deployment challenge, the goal is to make conclusions that generalize to the statistical population described by the problem and purpose statement, and, in particular, determine whether any of the submitted algorithms successfully solves the problem and purpose. Therefore the data sample should be representative of the statistical population, i.e., Figure \ref{fig2}.

Following a quantitative experimental design, a form of random data selection (\textbf{probability sampling}) should be used, such as stratified random sampling \cite{elfil2017sampling}; and the sample should be sufficiently large. As mentioned above, sampling of the training, test, and hold-out data should follow the same sampling method, and challenge organizers should make sure that the test and hold-out data are representative of the training data. For a deployment challenge, it is good practice to have a hold-out dataset with a \textbf{private leaderboard}. 

Ideally, an \textbf{hypothesis test} on the private leaderboard should determine whether an algorithm successfully solves the problem and purpose. To be able to assess whether a problem is solved, the truth criteria and metrics that are used should be representative of the problem as well. This is none-trivial and often requires various insight challenges and discussions with experts (both clinical and medical image analysis) in the field to reach a consensus. Multidisciplinary teamwork is essential in order to be able to design and set-up well designed deployment challenges. Metrics should be defined that fit clinical (practical) requirements \cite{nikolov18} and that could be used to test which algorithms would be good enough to be used in clinical practice (see Section \ref{sec4} for practical significance). Keeping these deployment challenges open (not time-bound) would allow for continuous monitoring (benchmarking) of state-of-the-art algorithms for specific problems, keeping track of how the field of research develops over time.

\subsection{Insight Design}\label{sec32}
For an insight challenge, the goal is not to generalize to the statistical population, but to gain insight into the performance of algorithms, given specific data, truth, and metrics. Insight challenges can provide small non-randomly selected data. When little data is available, it is more beneficial from a research perspective, to select a targeted sample, following a qualitative experimental design. For qualitative research it is common to purposefully select subjects, groups, or settings to  maximize understanding of the phenomenon under investigation \cite{Onwuegbuzie07}. 

In challenge design, you could think of gaining insight into the influence of various population variables (data, metrics, and ground truth) on algorithm results. Next to purposeful or purposive sampling, many variations of \textbf{non-probability sampling} exist. Like the definition of truth and metrics, the sampling method is closely linked to the purpose statement of a challenge. Some challenges have been set-up around a dataset that was already available (convenience sampling), in which case the purpose statement should clearly describe which insights are expected from this data, and truth criteria and metrics should be defined accordingly. 

Making existing data available through a challenge can be extremely beneficial, since it could lead to new insights. Also, if one group starts with opening up their data for a certain research problem, others might follow (snowball sampling), covering a larger part of the problem space (Figure \ref{fig1}). This might in turn inspire new \textbf{research directions} that focus on specific sub-parts of the problem, using for example maximum variation sampling, critical case sampling, extreme case sampling, or homogeneous sampling. Ultimately leading to a collaborative effort to set-up a deployment challenge that takes all insights from previous challenges into account.

Since insight challenges follow a qualitative experimental design, their leaderboard should be treated as a rough estimate of relative performance. Qualitative research is aimed at identifying underlying concepts or characteristics and aims to answer why a certain phenomenon may occur \cite{lune2016qualitative}. Therefore, leaderboard climbing and overfitting on the test data of the public leaderboard is less of a concern for insight challenges, since incremental improvements on insight challenges do not aid in solving the underlying problem. Further analysis of the results beyond the leaderboard is essential to gain insight into algorithm performance with respect to the data, truth criteria and metrics used in a challenge. Therefore, tools should be offered to be able to gain this insight and identify overfitting on the test data or leaderboard climbing. Most importantly, the leaderboard should be analyzed with respect to the qualitative design of the challenge and can therefore not be used to draw conclusions about generalization (Section \ref{sec4}).

\section{Applying the Framework: Significance, Generalizability, and Saturation}
\label{sec4}

In addition to a challenge's data, truth, and metrics defining its type, they also impact its ``efficacy''.  Others have covered the difficulties associated with selecting data \citep{Fan06}, truth \citep{Rodriguez12}, and metrics \citep{Fishbaugh17} to create ``effective'' challenges; however, much of that work has focused on maximizing a study’s statistical significance.  There are three difficulties with focusing on statistical significance when optimizing a challenge: 1) statistical significance is associated with testing a hypothesis and therefore is only relevant to quantitative experiments; 2) the complex interrelationship between data, metrics, and truth is oversimplified by commonly used measures of statistical significance; and 3) statistical significance is not a measure of practical significance or generalizability. In this section we provide a brief introduction to statistical significance and then introduce practical significance, generalization theory, and data saturation as important yet often overlooked concepts that impact a challenges “efficacy.”

Tests of statistical significance typically assume (1) that the test data is from a random sampling of the target population and (2) that the observed errors have a known, e.g., normal, distribution with respect to each independent variate \citep{arbuthnot1710argument}. The utility of these tests are degraded when these assumptions are incorrect, and in medical imaging, these assumptions are often incorrect: patients typically differ from one another with respect to a multitude of independent variables that are difficult to sample appropriately or that induce errors with non-normal distributions, e.g., with respect to age, gender, imaging device manufacturer.  Additionally, the definition of “significant” for any specific problem is highly sensitive to the (often arbitrarily) chosen values of $\alpha$, the sample size, and the use of one-tailed or two-tailed tests.

Practical significance, on the other hand, indicates if the findings have any decision-making utility. This implies a consideration of ``effect size'' \citep{Hojat04} as well as statistical significance. In the medical field, practical significance is often referred to as ``clinical significance'', and that term more clearly reflects the fact that small yet statistically significant differences between methods may have no impact on treatment or outcome. For example, a small yet statistically significant improvement in tumor boundary estimation may not enable a reduction in the radiation therapy treatment margins around that tumor.  

There are numerous tests for practical significance, such as Scoring-Guide Scales and Meaningful Reference Group Comparisons \citep{Greenland16}.  In scoring-guide scales, results from two methods are compared based on the percentage of individuals who are at each point on the scale, where that scale maps to the magnitude of clinical impact.  For example, scales can be defined, e.g., via retrospective clinical analyses or expert assessments, which map registration errors in radiation therapy treatment delivery to probability of tumor recurrence, and then those scales can be used to determine the clinical impact of one registration method versus another.  Somewhat similarly, in meaningful reference group comparisons, results from a method are compared to norms from chosen groups.  For examples, groups can be defined based on clinical expertise at tumor scoring, and the performance of a method can be equated to performing at the level of a novice, intermediate, or expert clinician. 

One of the strengths of measures of practical significance is that computed differences between competing methods are easy to understand.  For example, when using scoring-guide scales, conclusions are of the form, ``for these data, tumor recurrence rates would increase by 10\% if method Y is used''.  When using meaningful reference group comparisons, conclusions are of the form ``for these data, method Y performed at the level of an expert clinician''.  Additionally, many test of practical significance can be conducted even if the studies were qualitative, i.e., even if the test data was not representative of the complete clinical population.  Determining if the conclusions can be applied beyond the test data is accomplished via measures of generalizability. 

Generalizability is concerned with indicating what are the potential sources of the observed differences between the two methods being compared and indicating the amount of difference that can be a attributed to a source \citep{Li15}. Statistical significance tests may indicate that two diagnostic methods are not equivalent, but they will not inform the experimenter if their difference is driven by, for example, a patient's gender, by differences between the human experts who provided truth, or by the scanner on which some of the data was collected. 

In this paper, we take a broad definition of generalizability to also include the concepts of validity (the appropriateness of the data, truth, and metrics for the research question \citep{Leung2015}) and reproducibility (the ability to achieve consistent results under the described conditions \citep{Leung2015}). Admittedly, the exact definitions and metrics of generalizability, validity, and reproducibility are the topics of extensive research and debate. For example, one accepted practice defines validity as described above as ``internal validity'' and combines the concepts of reproducibility and generalizability as ``external validity'' \citep{Kukull12}.  We have chosen to combine these terms into a broad definition of generalizability, given that internal and external validity combine to impact how future researchers should interpret a challenge's published results.

Conceptually, testing for generalization is relatively straightforward.  Typically, analysis of variance (ANOVA) testing is used to partition observed variance into its primary components. The primary components can be separated into two categories: (1) those that arise from individual differences among the subjects and (2) those that arise from experimental ``facets'', e.g., differences between experimental factors such raters, imaging scanners, and such. By systematically accounting for variance, the ability of a study's results to be generalized to another case is revealed. \citep{reynolds98} \citep{Sinharay10}.   Also, by understanding a study's generalization limits (its biases and confounding facets), it is sometimes possible to adjust the study's results to reduce the impact of its facets and thereby improve the study's generalization capabilities \citep{Jagger08}.  

The widespread misinterpretation of statistical significance as a measure of practical significance or generalization has prompted statistical journals to discourage the use of tests of statistical significance \citep{wasserstein_asas_2016}. The Journal of Basic and Applied Social Psychology, for example, has gone so far as to ban the use of significance testing in its published papers. They instead require measures that focus on practical significance and generalizability \citep{Trafimow15}.  The trend away from statistical significance is also being prompted by the increased acceptance of the concept of data saturation, as explained next.

We define data saturation, in the context of challenges, as encompassing insufficiencies in the data, metrics, and/or truth that diminish the statistical and practical significance as well as the generalizability that can be achieved, even as sample size is increased.  Although the concept of data saturation has been hotly debated in the field of qualitative statistics \citep{Saunders18}, the prevailing premise is that a problem is suffering from data saturation when the utility of additional data is limited by the additional data’s redundancy with existing data.  These data redundancies may arise from limitations in sampling (e.g., purposeful versus random sampling), input detail (e.g., image resolution or contrast limitations make differentiation impossible), metric sensitivity and specificity (e.g., dice overlap is an incomplete measure of segmentation quality), and truth quality (e.g., expert assessments of breast density may be inconsistent).

Data saturation may be particularly relevant when trying to detect small yet statistically significant differences between algorithms using large databases, as is often the case with challenges and deep learning methods.  Consider, for example, the fact that a study has shown that it was possible to train a neural network to determine the source database of an image when using images from twelve popular, very large, ``real-world'' image databases \citep{Shankar17}.  This study suggests that there are biases in these databases, despite their massive sizes, that may have arisen from their samplings of ``nature,'' their acquisition protocols, or the class labels they provide. These biases will carry to systems trained and tested using them.  As a result, statistically significant differences measured with any one or combination of these data may not equate to differences in real-world performance (i.e., practical significance), and perhaps even more importantly, the source of errors that may impact real-world performance (i.e., generalization) will not be revealed by the measures of statistical significance. 

\begin{figure}[!t]
\centering
\includegraphics[scale=.5]{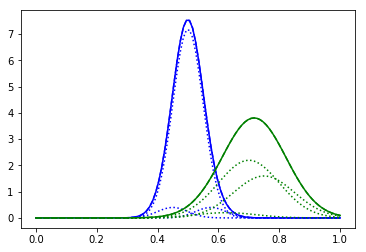}
\caption{\label{fig4} The two populations used in our study are mixtures of Gaussians.  Population A is shown in blue, and Population B is shown in green. Their component Gaussians are shown in dotted lines, and their combined distributions are shown in solid lines.  Population A approximates a normal distribution with p = 1e-5 using the Shapiro-Wilks test for normality.  Population B deviates from normal, with test for normality p = 3.1e-2.  These distributions were chosen to represent a problem that on the whole is well separated (offer a TPR of 0.93 at a FPR of 0.1)}
\end{figure}

Consider the toy example of a fictitious challenge whose problem is to distinguish between two different patient populations using a single feature.  In this one-dimensional feature space, each statistical population is a mixture of three Gaussians, see Fig. \ref{fig4}.  Such mixture distributions may arise, for example, from genetic variations in each population or from the progression of a disease.

Given 200 repeated random trials in which, for each trial, a random set of 200 training cases and a random set of 200 test cases were generated from altered versions of the statistical populations shown in \ref{fig4}.  A linear classifier was then defined by determining the threshold value that produces a 0.1 false-positive rate (FPR) on the training data. The 0.1 FPR target was chosen to reflect the common condition that an automated clinical system must avoid unnecessarily treating healthy subjects (i.e., false positives), while not drastically under-diagnosing those who have the disease. Using the estimated training data threshold, we then computed the FPR and true positive rate (TPR) for the test data.  Additionally, we generated a set of 200 test cases from the actual (un-altered) statistical populations and computed the threshold’s FPR and TPR values on that ``deployment'' data.  Using all 200 repeated runs, means and standard deviations of the altered data’s and deployment data’s testing FPRs and TPRs were recorded and were used to estimate the minimum statistically significant detectable difference at $\alpha$ = 0.01 and power = 0.95 using two-sided t-tests.  

\begin{figure}[!t]
\centering
\includegraphics[scale=.5]{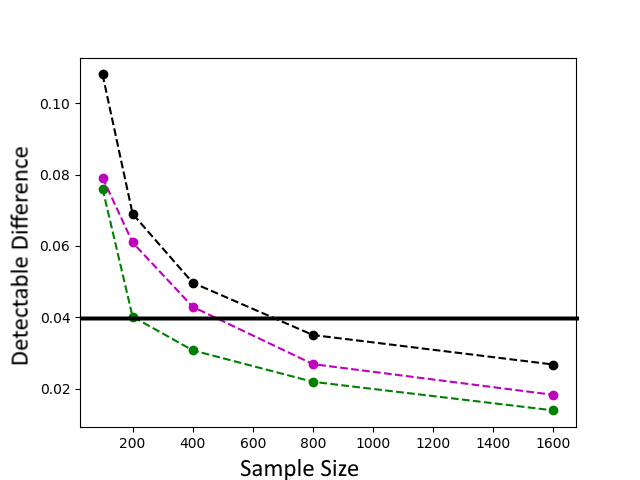}
\caption{\label{fig6}The impact of Data Augmentation (adding 2\% noise) and Mechanical Turk (mislabeling 2\% of the cases) for increasing sample size is particularly relevant to the study of data saturation.  The green (bottom) curve is the detectable difference curve when the actual statistical population data is used for training and test.  The black (top) curve occurs if the training data has had 2\% noise added to it.  The purple (middle) curve results when 2\% of the training cases are mislabeled.  The solid horizontal line highlights how additional noise and mislabelings can drastically increase the number of samples needed to achieve a given detectable difference when comparing methods.}
\end{figure}

In Fig. \ref{fig6} we illustrate the impact of Data Augmentation (adding 2\% noise to the training cases) and Mechanical Turk (increasing sample size, yet having 2\% of the training cases mislabeled).  While it is intuitive that these modifications resulted in non-optimal classifier design, in Fig. \ref{fig6} we show that these modification also drastically reduce the statistically significant differences that can be detected for a given sample size.  By mislabeling only 2\% of the data (e.g., 4 of 200 cases), instead of requiring 200 cases to achieve a 0.04 detectable difference between the performance of two classifiers, a challenge organizer would need to provide approximately 500 cases to achieve that same detectable difference, and if 2\% noise is added to the data, then over 700 cases would be needed.  

\section{Discussion \& Future perspective}
With this paper, we aim to make researchers aware of the value and complexity of challenges and their design and provide a framework to put challenges in perspective and determine next steps in challenge design. Most challenges organized in the field of medical image analysis thus far, are insight challenges. The first challenges often used a form of convenience sampling, i.e., evaluation data that happened to be available in the organizers' group (e.g., from another study) was used for organizing a challenge. This already led to new insights into algorithm performance on this specific dataset. It was a huge step forward in comparison to researchers writing papers claiming that their algorithm performed best on their private dataset \cite{price1986anything}, which did not give other researchers the opportunity to test their algorithm against the proposed algorithm. Later, a form of snowball sampling was used (i.e., groups, that knew each other or participated in a previous challenge, teamed up to share their data in a new challenge on the same topic), which led to insight into the performance of algorithms on data from different hospitals \cite{menze2014multimodal,bejnordi17,bakas2018identifying}.

What is currently lacking, is an overview of the problems that researchers in the field of medical image analysis are working on and how challenges relate to these research problems. The paper "Grand Challenges in Biomedical Image Analysis" \cite{Ginneken18} is a excellent first step in providing an overview; however the quality of various challenges discussed in that paper is unclear, e.g., which part of the problem space (Figure \ref{fig1}) each challenge covers and what insights can be gained from these challenges. Without these details, it is hard to determine new research directions that would drive the field forward. 

The medical image analysis field needs tools that help guide researchers in proper challenge design and that generate an overview of the state-of-the-art for each problem. Ideally, multi-skilled teams should be organized that include web developers, research software engineers, researchers, statisticians and clinicians to join forces and set-up well-designed challenges that boost the field. Developing these tools and accomplishing these goals might be significantly accelerated by changes to the reward system used by funding agencies and journal article reviewers.

Whereas insight challenges are very interesting from a research perspective, deployment challenges are perhaps particularly well suited to be set-up by companies, as part of their clinical trials or cohort studies. This could aid in bridging the gap between industry and academia. If data, truth criteria, and metrics are representative of the problem, direct assessment of algorithms from academia could result in more practical and accelerated use of academic algorithms. Some examples of first steps in this direction can be found on Kaggle (kaggle.com). Further research is required, in particular, to determine metrics that capture practical significance, e.g., aid in determining when an algorithm is good enough to be used in clinical practice?. For some problems, multiple insight challenges with targeted samples (purposive samples) should be set-up in order to better understand the statistical population variables that influence algorithm performance (e.g. generalization). Examples of this are extreme case sampling (e.g., determining what is the influence of extreme cases of pathology or artefacts?), critical case sampling (e.g., selecting forms of pathology that are clinically ambiguous but critical), or maximum variation sampling (e.g., maximizing acquisition variation or variation in pathology).  

The focus of challenges should be on solving or gaining insight into the challenge's target problem and purpose. Analysis of algorithm results beyond the leaderboard will aid in identifying leaderboard climbing, but future research should focus on finding better algorithms to detect or combat leaderboard climbing. Future research should also focus on the quality of the ground truth and how to assess whether algorithms outperform humans. Interesting examples are letting a panel of experts perform the same task as the algorithm with time limitations \cite{bejnordi17} or presenting the false positives of the top performing algorithms to experts for re-evaluation \cite{veta2015assessment}.  

Furthermore, when designing a challenge, it is important to not only focus on statistical significance and sample size.  Practical significance and generalization must also be computed and reported, and therefore data saturation concerns must also be monitored. In particular for challenges, increasing sample size at the cost of adding noise or accidentally mislabeling even a small portion of the data can have drastic detrimental effects.   This is clearly demonstrated for the simple single-feature, two-class experiments presented in this paper, and the potential importance of these lessons only increases for problems with higher-dimensional feature spaces and complex decision boundaries, e.g., the types of problems that are now being investigated using deep learning and other data-drive machine learning methods.  The measured performance of an algorithm on an insight challenge may appear better or worse than the performance that would be achieved if that algorithm was deployed. Measures of statistical significant on challenges with a qualitative design will not indicate if an algorithm’s performance on a challenge is close to or far from what its performance would be if deployed. Aspects of these shortcomings are also discussed in other publications in the machine learning field. For example, Masko et al. \citep{Masko15} explored the detrimental effects on generalization when class priors are altered in a neural network’s training data.

Finally, more advanced infrastructures and additional funds are required to support open challenges (not time-bound) and submission of algorithms instead of algorithm results. In other words, enable continuous monitoring of state-of-the-art algorithms for various research problems (benchmarking), keeping track of how the field develops over time. Dynamic benchmarks, where data can be added after quality checks and algorithms can be re-run is an interesting future prospect to quickly track progress. An example of providing evaluation as a service for medical image registration is described in \cite{marstal2019continuous}. Although many great initiatives exist, such as dream challenges (dreamchallenges.org), Sage bionetworks (sagebionetworks.org), codalab (codalab.org), COMIC (grand-challenge.org/Create\_your\_own\_challenge) and Kaggle, it remains a challenge to keep these platforms sustainable and to track progress in the field.

\section{Conclusion}
The challenge framework presented in this paper offers a unified approach to the design of qualitative and quantitative challenges, i.e., insight-generating and problem-solving challenges.  Both types of challenges are beneficial to research, but the design decisions that define a challenge must consistently hold to the challenge’s objective of being qualitative or quantitative.  Furthermore, when data augmentation, mechanical turk, novice raters, and/or simulated data is involved; care must be taken to ensure that their detrimental effects are closely assessed.  Additionally, assessing statistically significant differences on leaderboards should not be the end goal of a challenge.  Deployment challenges should have hold-out datasets on which generalization and practical significance are measured, and insight challenges should conduct qualitative analyses to determine fruitful research directions and to design effective future challenges.

\section{Acknowledgments}
The authors would like to acknowledge Bram van Ginneken and Tobias Heimann, who have made significant contributions to the concept of challenges in the medical image analysis field, and to all challenge organizers who invested their time in open science to advance this concept. For dr. A.M. Mendrik, this work was funded, in part, by IMDI Grant 104002002 (Brainbox) from ZonMw, the Netherlands Organisation for Health Research and Development, and in part, by the Netherlands eScience Center, Research and Development budget. For dr. S.R. Aylward, this work was funded, in part, by National Institute Of General Medical Sciences (NIGMS) and the National Institute Of Biomedical Imaging And Bioengineering (NIBIB) via NIH grant R01EB021396 and by the National Institute of Neurological Disorders and Stroke via NIH grant R42NS086295.

\bibliographystyle{unsrt}  

\end{document}